# Spontaneous particle desorption and "Gorgon" drop formation from particle-armored oil drops upon cooling

Diana Cholakova,[1] Zhulieta Valkova,[1] Slavka Tcholakova,[1*] Nikolai Denkov,[1] and Bernard P. Binks[2]

[1]*Department of Chemical and Pharmaceutical Engineering*
*Faculty of Chemistry and Pharmacy, Sofia University,*
*1 James Bourchier Avenue, 1164 Sofia, Bulgaria*
[2] *Department of Chemistry and Biochemistry, University of Hull, Hull HU6 7RX, UK*

**\*Corresponding author:**
Prof. Slavka Tcholakova
E-mail: sc@lcpe.uni-sofia.bg
1 James Bourchier Avenue,
1164 Sofia,
Bulgaria
Tel: +35928161698
Fax: +35929625643

**Keywords:** Pickering emulsion, particle desorption, self-shaping drops, rotator phases, non-spherical drops

**ABSTRACT**

Hypothesis

Drop "self-shaping" is a phenomenon in which cooled oily emulsion drops undergo a spectacular series of shape transformations (Denkov *et al*., *Nature* **528,** 2015, 392). Solid particles adsorbed on the oil-water interface could affect this drop self-shaping process in multiple ways which have not been studied.

Experiments

We prepared Pickering emulsions stabilized by spherical latex particles and afterwards added surfactant of low concentration which enabled drop self-shaping. Next we observed by optical microscopy the processes which occur upon emulsion cooling.

Findings

Several new processes were observed: (1) Adsorbed latex particles rearranged into regular hexagonal lattices upon freezing of the surfactant adsorption layer. (2) Spontaneous particle desorption from the drop surface was observed at a certain temperature – this phenomenon is rather remarkable, as the solid particles are known to irreversibly adsorb on fluid interfaces. (3) Very strongly adhered particles to drop surfaces acted as a template to enable the formation of tens to hundreds of semi-liquid fibers, growing outwards from the drop surface, thus creating a shape resembling the Gorgon head from Greek mythology. We provide mechanistic explanations of all observed phenomena using our understanding of the rotator phase formation on the surface of cooled drops.

# INTRODUCTION

Emulsions are widely used in the food industry, pharmacy, cosmetics, drug delivery, home-and-personal care and many other industries.[1-3] Usually, they are prepared in the presence of low-molar-mass surfactants or macromolecules which adsorb on the oil-water interface and stabilize the drops to coalescence. The possibility for stabilization of emulsions without surfactants, only with solid particles adsorbed on the oil-water interface, has been known for more than a century.[4] Research interest in these particle-stabilized ("Pickering" or "Ramsden") emulsions has sparked in the last decades,[5-11] due to specific advantages over the more common surfactant-stabilized emulsions[11]: (1) The particle-stabilized emulsions have high stability to drop coalescence and drop Ostwald ripening; (2) they are surfactant-free which may be a major benefit in some applications, such as cosmetics and pharmaceutics; (3) the adsorbed particles may be modified and functionalized in different ways, thus bringing additional useful features to the final emulsions,[12] such as desired electrical and magnetic properties, responsive behaviour *etc*.

The stability of the particle-stabilized emulsions is governed by the high desorption energy needed to be overcome in order to detach a particle adsorbed at an interface[5,7,13]:

$$E_{des} = \pi \sigma R^2 \left(1 \pm \cos\theta\right)^2, \tag{1}$$

where σ is the interfacial tension between the two immiscible fluids, $R$ is the radius of the particle and $\theta$ is the three-phase contact angle between the fluid interface and the surface of the solid particle. The sign inside the brackets depends . Even for relatively small particles of nanometer size, this energy is of the order of thousands $k_B T$ (thermal energy) and for particles with size 1 μm it becomes more than $10^7$ $k_B T$. Therefore, usually the adsorption of particles on the liquid-liquid interface is considered to be irreversible, which prevents their recycling. There are many applications, however, where such recovery of the particles attached to the interface may be desirable, or alternatively the emulsions may need to be destroyed.[14-16] For example, emulsions formed during the petroleum extraction from oil sands are difficult to break, because they are often stabilized by fine solids such as asphaltene aggregates, wax particles, silica and clay minerals. These emulsions should be demulsified in order to allow subsequent petroleum usage.[17] Particle separation and emulsion destabilization is also desired in nanoparticle catalyst systems.[18,19]

Several different mechanisms have been proposed for the destabilization of Pickering emulsions.[14,16,18-28] Most of them rely on the use of stimuli-responsive particles[18-26]. Usually, the mechanisms governing the removal of the stimuli-responsive particles from interfaces involve either a change in the particle or an external force, i.e., alteration of the particle wettability[12] in response to the change in some external factor like pH[12,18,19], salt concentration[20,21], presence of electric[12,22] or magnetic field[23,24,29], microwaves[25] or light[26]. Particle detachment from fluid interfaces is also observed if mechanical (*e.g.* shear) forces are applied to the system.[15,27,30] Usage of periodic compression-expansion due to an ultrasound wave also causes detachment of solid particles from a bubble surface.[15] Gold nanoparticles detach from the water-octafluoropentylacrylate interface upon compression of a pendant drop.[27] Particle detachment in Pickering emulsions is also observed sometimes when appropriate surfactant is added to the emulsion.[28] None of these mechanisms are widely applied or system-independent and new general mechanisms are needed both for practical industrial purposes and to understand the difficult fundamental problems of how to desorb almost irreversibly adsorbed particles from an interface.

In the current study we demonstrate a new mechanism for particle detachment from emulsion drops governed by the formation of a plastic rotator phase beneath the drop surface upon cooling. Rotator phases are intermediate phases between the isotropic liquid phase and a crystalline phase. Molecules do have long-range positional order with respect to their translational degree of freedom, but unlike in crystals where rotation is also fixed, the molecules have a full or partial rotational degree of freedom around their long axis.[31-35] Rotator phases are known to form for long-chain alkanes, alkenes, alcohols, asymmetric alkanes and other organic molecules and their mixtures.[35] These phases are very similar to the so called α-phases in triglycerides.[36]

Rotator phases are observed for bulk substances, on the alkane-air interface and also in confinements such as nanoporous materials, polymeric microcapsules and emulsion droplets.[35] Recently, we showed that formation of rotator phases upon cooling of oil-in-water (o/w) emulsion drops stabilized by long-chain length surfactants leads to formation of various non-spherical fluid particles such as hexagonal platelets, triangular and tetragonal platelets and rods.[37,38] These shapes can be preserved also upon complete freezing of the oil droplets. Furthermore, the self-shaping phenomenon was related to a self-emulsification process in which drops spontaneously burst into multiple smaller droplets only due to cooling and heating without any mechanical energy input.[39,40] A drop size decrease from 33 μm to well below 1 μm was observed for the most efficient systems.

The mechanism of the drop self-shaping phenomenon originally proposed by Denkov *et al.*[37,38] is illustrated schematically in Figure 1 and can be explained as follows: when the surfactant adsorption layer consists of molecules with not-too-voluminous head groups and tails with chain length comparable to the length of the alkane molecules, upon cooling the surfactant molecules arrange close enough so that their hydrophobic tails eventually freeze and form a frozen adsorption monolayer which may contain also intercalated alkane molecules. This frozen monolayer serves as a template for the ordering of the alkane molecules from the inner layers, which arrange into multilayers of plastic rotator phase. The enthalpy gain from the molecular ordering compensates for the interfacial energy increase due to the significant drop area increase upon the self-shaping transformations.

An alternative mechanism of the process was proposed by Guttman *et al.*[41-43] These authors explained the observed deformations for their system (hexadecane $C_{16}$ drops dispersed in $C_{18}$TAB surfactant solution) with significant decrease of the interfacial tension (IFT) which becomes ultra-low and even transiently negative. However, the interfacial tension and differential scanning calorimetry (DSC) measurements made by Denkov *et al.*[44,45] showed that this is not the case for the many systems studied by them, including ones in the present paper, which include not only a single "ultra-pure" alkane-surfactant system but various alkanes and other oils combined with many different surface-active compounds.

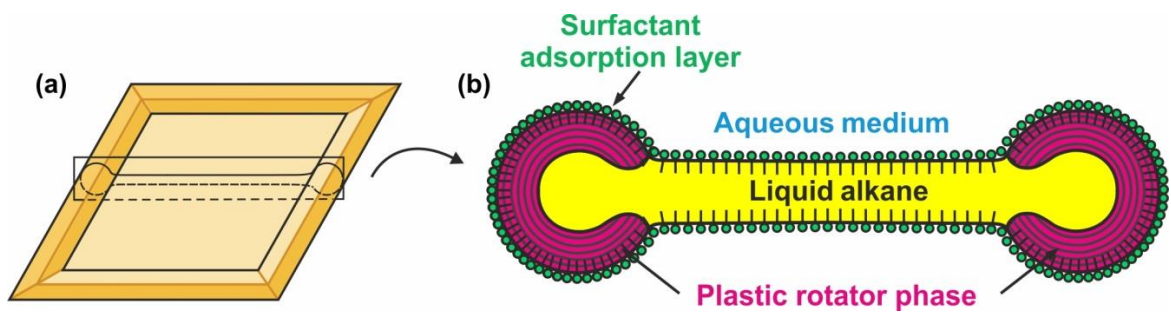

**Figure 1.** Schematic presentation of the mechanism of the drop self-shaping process: (a) Tetragonal platelet with thicker frame at the particle periphery. (b) Cross-section of the platelet with plastic rotator phase at the platelet periphery (in magenta) and liquid interior (in yellow). The surfactant adsorption layer which freezes and serves as a template for the ordering of the alkane molecules next to the drop surface is shown in green and black.

In the current study we demonstrate how the drop self-shaping process changes in the presence of adsorbed particles. We show that depending on the surfactant, electrolyte and particles type used for emulsion formation, the particle-particle interactions are altered and particle detachment upon cooling for mobile particles and formation of new shapes resembling "Gorgon head" for jammed particles is observed.

**EXPERIMENTAL**

*Materials*

As dispersed phases in the prepared emulsions we used *n*-alkanes with different chain lengths, denoted in the text as $C_n$, where *n* varied between 14 and 16 C-atoms. For emulsion stabilization we used two classes of nonionic surfactant – polyoxyethylene alkyl ethers, $C_nEO_m$ with *n* varied between 16 and 18, and *m* varied between 2 and 20, and polyoxyethylene sorbitan monoalkylates, $C_nSorbEO_{20}$ with *n* varied between 16 and 18. Hexadecyl trimethylammonium bromide ionic surfactant $C_{16}TAB$ was also tested. Detailed information of the chemical purity, physical properties and producers of the alkanes and surfactants used is presented in Supporting Information Tables S1 and S2.

Four types of surfactant-free latex particles were tested. Carboxyl latex (4% w/v), chloromethyl latex (4% w/v) and sulfate latex (8% w/v) with average particle diameter of 1 μm were purchased from Thermo Fisher Scientific. Chloromethyl latex particles with average diameter of 1.6 μm were purchased from IDC spheres. All aqueous solutions were prepared with deionized water, which was purified by an Elix 3 module (Millipore).

Sodium chloride with purity 99.8% was purchased from Sigma and calcium dichloride dihydrate, purity 99%, was purchased from ChemLab. The electrolyte concentration was varied between 100 mM and 500 mM for NaCl and between 0.5 and 10 mM for $CaCl_2$. The electrolyte is added to overcome the electrostatic barrier for particle adsorption on the oil-water interface as we have shown in our previous work.[46]

Most of the substances were used as received without further treatment. In part of the experiments (indicated in the text) we used hexadecane, which was purified from surface-active contaminations by passing through a glass column filled with Florisil adsorbent. All processes described in the paper are observed with both purified and non-purifed hexadecane. The only difference is small deviations in the temperatures at which these processes are observed, as explained in the result section.

*Emulsion preparation*

The emulsions were prepared according to the following procedure: first the alkane was added into the already prepared aqueous electrolyte solution without surfactant. Then the latex particles were added to the mixture (particle concentration 0.05 wt.% with respect to the whole emulsion). The obtained electrolyte solution-alkane-particle mixture was emulsified using a rotor-stator Ultra-Turrax IKA T25 homogenized for 3 min at 8000 or 13,500 rpm. After emulsification, a 10 mM aqueous surfactant solution was added to the emulsion by gentle hand shaking. The final surfactant concentration (calculated accounting for the dilution) was varied between 1 and 3 mM. The addition of surfactant enabled the spontaneous drop self-shaping deformations. In the final emulsions, a mixed particle-surfactant adsorption layer was formed.

*Optical observations upon cooling*

A sample of the prepared emulsion was placed in a glass capillary with a rectangular cross-section: 50 mm length, 1 or 2 mm width and 0.1 mm height. The capillary was enclosed within a custom-made cooling chamber with optical windows for microscopy observations, see Supporting Information Figure S1. The observations were performed with optical microscopes Axioplan and AxioImager.M2m (Zeiss, Germany) in transmitted white light or cross-polarized white light. Long-focus objectives ×20, ×50 or ×100, combined with built-in cross-polarizing accessories of the microscope, were used to observe the drops upon sample cooling. In most of the experiments we used an additional λ plate (compensator plate) which was situated between the polarizer and the analyzer, the latter two being oriented at 90° with respect to each other. The λ plate was oriented at 45° with respect to both the polarizer and the analyzer. Under these conditions the liquid background and the fluid objects have typical magenta color, whereas the birefringent areas appear brighter and may have intense colors.[38]

The temperature of the cooling chamber was controlled by a cryo-thermostat (Julabo CF30) and measured close to the emulsion location using a calibrated thermo-couple probe, with an accuracy of ± 0.2 °C. Experiments were carried out at different cooling rates, varied from 0.1°C/min to 3°C /min.

*Interfacial tension measurements*

The hexadecane-water interfacial tension, σ, was measured by drop-shape analysis as a function of temperature.[44,47,48] The shape of millimeter-sized pendant oil drops, immersed in the aqueous surfactant solution, was recorded and analyzed by the Laplace equation of capillarity to determine the oil-water interfacial tension (instrument Krüss DSA100, Germany). The thermostating cell TC40 was used to vary the temperature of the measured system with a precision of ± 0.2 °C. The time required for cooling a drop of millimeter size was estimated as τ ≈ 10 s.[44] In a typical experiment, the temperature was decreased with a rate mimicking that in the actual experiments with slowly cooled emulsions (0.1°C/min) allowing us to compare the data of the two types of measurements. In separate experiments, the actual temperature at the position of the pending drop was measured by a calibrated thermocouple.

*Zeta potential measurements*

The zeta potential measurements of the latex particles dispersed in various aqueous solutions (different surfactant types and concentrations) and of the emulsion samples were conducted with a Zetasizer Nano-ZS instrument (Malvern Instruments Ltd,. Malvern, UK), equipped with a high electrolyte concentration Zeta potential cell (ZEN1010). The cell is designed to measure the zeta potential in high electrolyte concentration samples of small volume. Measurements were performed at different temperatures between 10 and 25 °C. At least 10 independent measurements of each sample were made in order to ensure reproducibility of the results.

**RESULTS AND DISCUSSION**

Here we consecutively describe and discuss the phenomena observed upon cooling of emulsion droplets stabilized by a combination of surfactant and adsorbed particles.

*Particle rearrangement into packed structures on the drop surface, driven by local freezing of surfactant adsorption layer*

At temperatures well above the bulk melting temperature of the dispersed oil, the adsorbed particles on the drop surface are evenly distributed over it, see Figure 2a,e. Upon cooling two types of particle rearrangement are observed. In some cases, all of the particles come close to one another, forming an ordered particle monolayer on the drop surface, Figure 2b-d, i-k. In other cases for similar coverages, the particles rearrange into a mesh of particle-packed strips around wide circular areas without particles, Figure 2f,l. The rearrangement process is observed with all tested particle-surfactant systems where (i) the particle coverage of the emulsion drop is only partial and (ii) formation of rotator phases upon cooling is known to occur.[38] The differences between these two cases are due to the different number of nucleation points for the surface freezing driving the particle rearrangement.

The observed rearrangement process is explained with an interfacial tension gradient along the surface, emerging due to the freezing of parts of the surfactant adsorption layer, see Figure 3. The freezing is not instantaneous on the entire surface but happens after the formation of one or several nuclei on the drop surface. Freezing lowers locally the interfacial tension at the frozen areas whereas it remains higher in the areas in which surfactant tails are still fluid. The interfacial tension gradient on the two sides of the particle leads to a net positive force in the direction of increase of interfacial tension and particles move in that direction. When the initial nucleus formation is relatively difficult, the surface starts to freeze from a single location and all particles rearrange to a single area on the drop surface (Figure 2b-d), while when several nuclei emerge simultaneously, the particles rearrange in the regions between the different frozen areas (Figure 2f,l).

An intermediate case occurs when initially few main nuclei are formed resulting in particle arrangement on the drop surface squeezed into a triangular shape between the expanding empty regions, often with a "hole" in the particle array. Usually, upon further cooling the inner "defect" collapses and the triangular layer of particles also becomes close-packed. The rearrangement is likely due to the building pressure from the forming stronger plastic phase areas on the sides of the triangle region, see Figure 2i-k. We note that although we use the term "surfactant/tails freezing", the molecular order of the hydrophobic chains of the surfactant molecules is not a perfect crystal and was found to have very similar characteristics (*i.e.* molecular cross-section and layer thickness) to the one found in the rotator phases forming in the respective bulk alkane.[49-52]

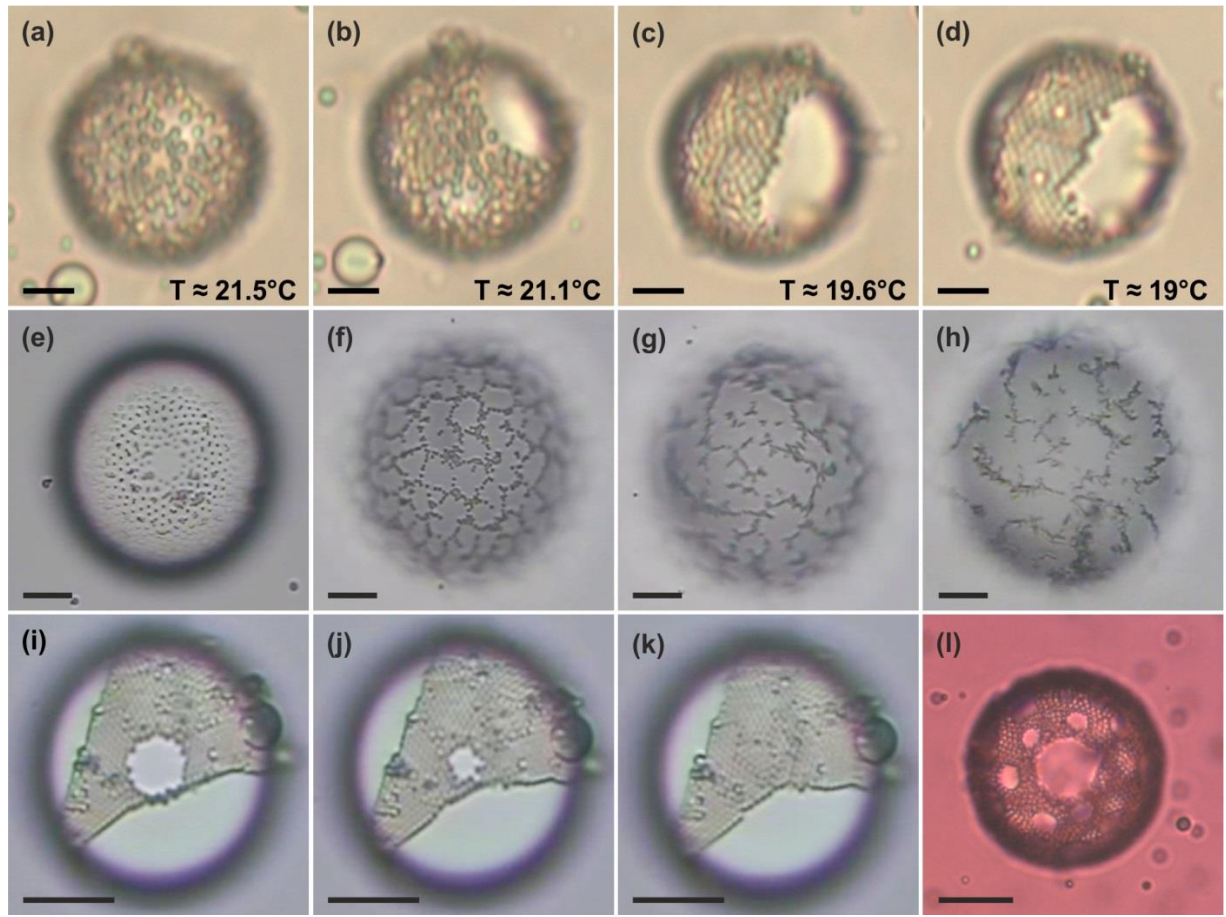

**Figure 2. Particle rearrangement on drop surface observed upon cooling.** (a-d) Particles rearrange on the left side of the droplet. This is most probably a result of the formation of a single crystalline nucleus on the drop surface. (e-f) Particle arrangement into a network. This behaviour is expected if many different crystalline nuclei are formed approximately at the same time. After the initial formation of a uniform network, it is disturbed upon further cooling (g,h). (i-k) Particles have arranged into a "triangular" monolayer with a circular "defect" inside it. Upon cooling, this empty patch collapses and the particles rearrange to form a close-packed monolayer. (l) Ordered particles with several circular "defect" areas without particles inside them. The systems are: (a-d) Hexadecane droplet with adsorbed sulfate latex particles; aqueous solution contains 1 mM $C_{18}EO_{20}$ and 100 mM NaCl. (e-h) Pentadecane droplet with adsorbed carboxyl latex particles; aqueous solution contains 1 mM $C_{18}EO_{20}$ and 7.8 mM $CaCl_2$. (i-l) Hexadecane droplet with adsorbed chloromethyl latex particles of 1 μm diameter; aqueous solution contains 1 mM $C_{18}EO_{20}$ and 100 mM NaCl. In all experiments non-purified oils are used. Scale bars: (a-d) 5 μm; (e-l) 20 μm.

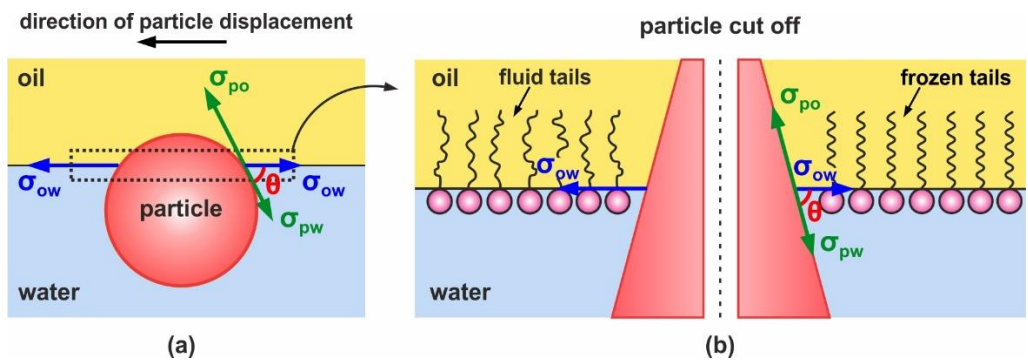

**Figure 3. Mechanism of particle rearrangement at emulsion drop surface.** (a) Upon cooling, particle displacement along the drop surface is observed. It is attributed to the emergence of a surface tension gradient on the two sides of the adsorbed particle. (b) Zoom-in of the oil-water interfacial area. In this scale, the particle edges look like straight lines because molecules are about two orders of magnitude smaller than the particle. Upon cooling, surfactant tails eventually freeze after the formation of a surface nucleus. The nucleus does not form instantaneously over the whole drop surface, therefore an interfacial tension gradient emerges. In the regions with frozen tails the interfacial tension decreases, whereas in the regions where the tails are still fluid the interfacial tension is higher. The interfacial tension gradient results in a net positive force in the direction of increase of the interfacial tension.

The particle rearrangement due to the freezing of the surfactant adsorption layer allow us for the first time to relate the results obtained for the interfacial tension in the system measured by the drop shape analysis method and the drop shape transformations observed. Illustrative results for the measured interfacial tensions for the systems containing purified and non-purified $C_{16}$ drops in solutions containing 3 mM $C_{18}EO_{20}$ or 3 mM $C_{16}SorbEO_{20}$ and 100 mM NaCl are presented in Figure 4a and Supplementary Figure S2. For the $C_{18}EO_{20}$ system with purified $C_{16}$, the interfacial tension remains almost constant down to 20 °C when it starts to decrease due to the freezing of the surfactant adsorption layer. In the experiments with purified oily drops (performed with the exact same temperature protocol as the one used in the IFT measurements), the particle rearrangement starts at the same temperature, $20 \pm 0.1$ °C, Figure 4b. When these experiments are repeated with non-purified oil, the same process is observed, however the surface freezing temperature is slightly higher and respectively the particle rearrangement begins at higher temperature, Figure 4a,c. A similar trend is also observed for the $C_{16}SorbEO_{20}$ system, see Supplementary Figures S2, S3a-d and S4a-d. In this case, again the slope of the dependence of IFT *vs* T changes at the temperatures at which the latex particle rearrangement is observed indicating changes in the surfactant adsorption layer.

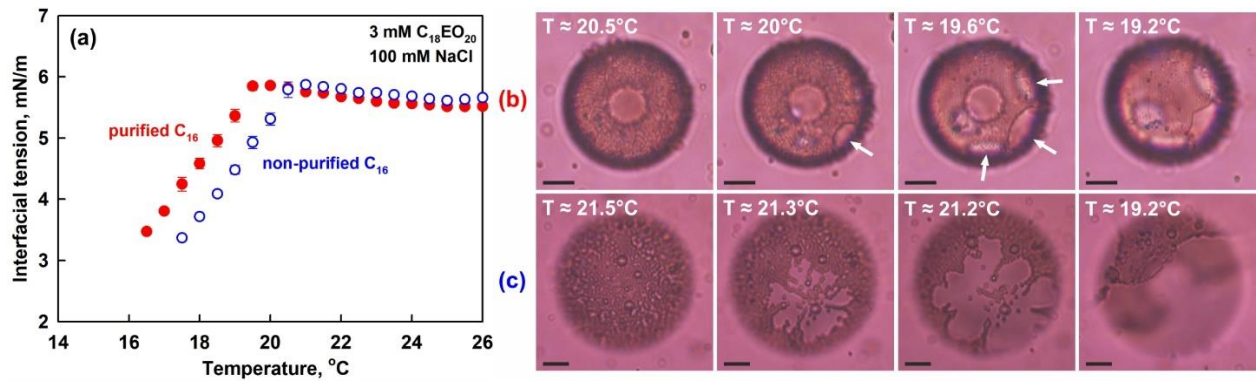

**Figure 4. Interfacial tension and its relation to particle arrangement.** (a) Interfacial tension measured by drop shape analysis method as a function of temperature for purified (full symbols) and non-purified (empty symbols) hexadecane droplet immersed in aqueous solution containing 3 mM $C_{18}EO_{20}$ and 100 mM NaCl. (b-c) Particle rearrangement observed with (b) purified $C_{16}$ and (c) non-purified $C_{16}$. Both emulsions are prepared with 3 mM $C_{18}EO_{20}$, 100 mM NaCl and 1 μm chloromethyl latex particles. The particle rearrangement process is observed in both systems, however the temperatures at which the process starts deviate slightly: 20 °C for purified oil and 21.3 °C for non-purified oil. The same trend is observed also in the interfacial tension measurements shown in (a). The observed process is caused by the freezing of the surfactant adsorption layer. All experiments are performed under equivalent temperature protocol (0.1 °C/min cooling rate). Scale bars: 10 μm.

Freezing of the surfactant adsorption layer leads to formation of a dense particle monolayer on part of the drop surface, while the rest of the surface becomes free of particles, Figure 5. At the level of the particle equators we have a close-packed 2D-array of solid particles, Figure 5c. However, if we consider now the same particle array at the level of the oil-water interface, the picture is quite different, Figure 5b. The studied particles are more hydrophilic and therefore they are preferentially immersed into the aqueous phase, *i.e.* the three-phase contact angle particle-oil-surfactant solution (measured through the aqueous phase) is smaller than 90°, Figure 3. This means that the cross-section between an adsorbed particle and the oil drop surface is a circle with diameter smaller than the particle diameter, Figure 5b.

After the initial surfactant tails freezing leading to particle rearrangement on the drop surface, upon subsequent cooling or after some time, the whole surfactant adsorption layer freezes. Therefore, the rearranged particles become "trapped" at their places being surrounded by a frozen adsorption layer. This is observed in the experiments as particles become less mobile and almost do not move over the drop surface, even when discrete or in small clusters not closely packed with others.

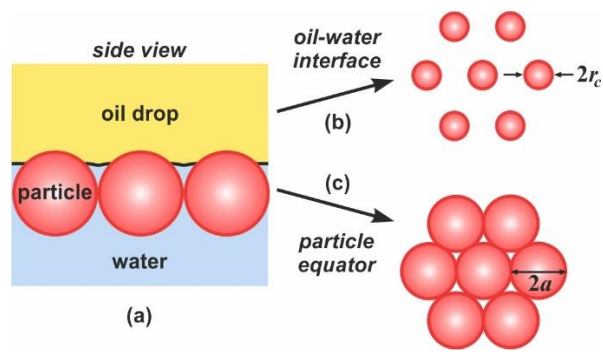

**Figure 5. Formation of dense particle monolayer.** (a) Side-view of the particle monolayer. (b) Oil drop surface as viewed from the oil-surfactant solution interface. The latex particles occupy only part of the drop surface because their contact angle is smaller than 90°. (c) Oil drop surface as viewed from a plane above the adsorbed particles. The number of particles in the monolayer is restricted with respect to the geometrical dimensions of the particles; the smallest distance between the particles can be so that they touch each other in their equators.

The next stage of the process is formation of the rotator phase multilayers close to the drop surface which lead to the drop self-shaping process. Three main types of behaviour were observed at this stage, depending on the specific system: (1) particle detachment from the drop surface, (2) Gorgon drop formation with multiple fibers extruding though the gaps between the adsorbed particles from the drop surface, and (3) drop shape deformations in the presence of particles. In the next section we discuss in sequence each of these processes.

*Particle behaviour upon rotator phase formation*
  *(a) Particle detachment upon cooling*
A non-trivial process of particle detachment from the oil droplet surface was observed in several of the studied emulsion systems, see Figure 6, Supplementary Movies S1 and S2 and Supplementary Figures S3e-h and S4e-h. The particle detachment process begins at almost the same moment when the drop shape deformations begin for drops of the same system but without adsorbed particles, *i.e.* when the multilayers of plastic rotator phase start to form on the drop surface. However, the presence of adsorbed particles hinders the regular rotator phase formation and suppresses the usual course of the drop shape deformations. Therefore, typically some fraction of the particles desorbs first and only afterwards the oil drops start to deform, see Figure 6 for example.

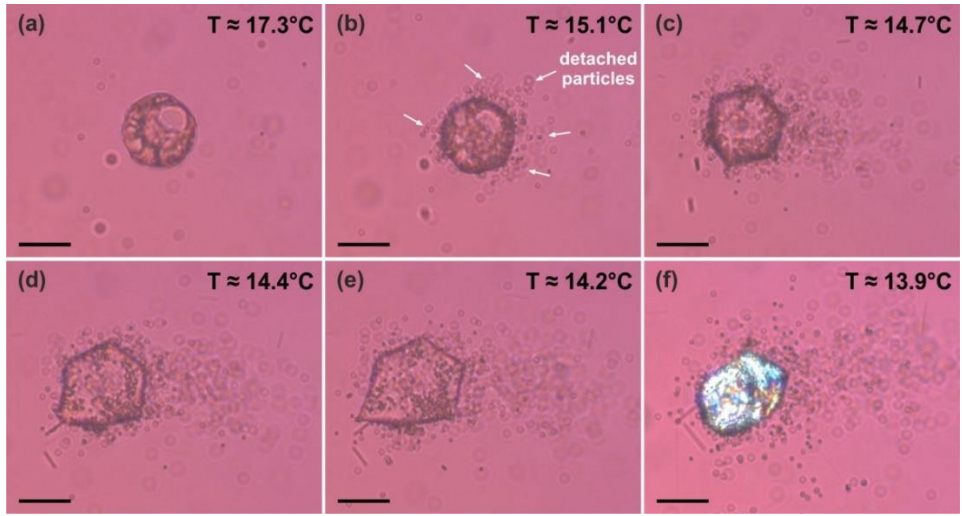

**Figure 6. Chloromethyl latex particle detachment observed upon cooling.** (a) Initially, the particles are ordered on the oil drop surface. (b) Upon cooling, the drop starts to change its shape and part of the adsorbed particles detach from the drop surface. (c-e) The particle detachment continues until the final drop freezes. (f) Frozen drop. The particles observed around it have detached during the cooling process. The system is: non-purified hexadecane droplet in 100 mM NaCl solution containing 3 mM $C_{18}EO_{20}$ surfactant. Cooling rate 0.5 °C/min. Scale bars: 20 μm.

The particle detachment process proceeds as follows: a particle situated at the edges of an area without particles escapes from the ordered monolayer and afterwards the particle detaches from the drop surface. The fact that the particles first escape from the ordered monolayer and only afterwards they detach from the surface is related to the friction between the particles. A particle surrounded by other particles would need to overcome also the friction with the surrounding particles in order to detach from the surface, whereas the particles situated in the layer periphery can escape first and then detach from the drop surface without the additional energy required to overcome the friction forces.

To understand the conditions for particle detachment one should analyze theoretically the changes in the surface energies upon emulsion cooling. One can start this analysis with the classical equation of Young-Laplace:

$$\cos\theta_i = \frac{\sigma_{po} - \sigma_{pw}}{\sigma_{ow}} \qquad (2)$$

which relates the three-phase contact angle at the particle surface, $\theta_i$, with the three interfacial tensions where "$p$" denotes the particle, "$w$" the water and "$o$" the oil phase. According to eq. (2), significant change in the three-phase contact angle could be achieved only upon changes in the interfacial tensions. Indeed, in the system for which the particle detachment process is most pronounced, chloromethyl latex particles adsorbed onto alkane drops in the presence of $C_{18}EO_{20}$ surfactant solution containing NaCl or $CaCl_2$, decrease in $\sigma_{ow}$ with decreasing temperature is observed experimentally, Figure 4a. This decrease may lead to a decrease in the three-phase contact angle and particle detachment.

However, there are several reasons which lead us to expect that the process of particle detachment is more complex in the systems studied here. First, for most emulsions we do not measure significant changes in $\sigma_{ow}$ that could explain the required change of the contact angle according to eq. (2). An example for such a system is $C_{16}/C_{16}SorbEO_{20}$ in the presence of NaCl with adsorbed chloromethyl latex particles where extensive particle detachment is observed. In this system, the oil-water interfacial tension has a very small change in the slope at the moment when the particles begin to order, thus indicating the onset of the adsorption layer freezing, but the changes in the value of $\sigma_{ow}$ are so small that they could not explain the observed particle detachment; see Supplementary Figures S2, S3 and S4.

We measured also the zeta potential of the chloromethyl particles dispersed in salt-surfactant solution to check whether $\sigma_{pw}$ changes significantly upon cooling, because this interfacial tension also affects the contact angle $\theta_i$. The measured zeta potential for chloromethyl latex particles in 100 mM NaCl (without surfactant) was $\zeta \approx -53.8 \pm 1.3$ mV, whereas in the presence of 100 mM NaCl and 3 mM surfactant it was $\zeta \approx -9.2 \pm 1.4$ mV for $C_{18}EO_{20}$ and $\zeta \approx -11.3 \pm 0.7$ mV for $C_{16}SorbEO_{20}$. The measured values between 10 °C and 25 °C were the same in the frame of the experimental error for both systems and no dependence on the temperature was observed. These experimental results show clearly that the surfactant molecules adsorb on the surface of the latex particles, screening the surface charge, but no changes are detected upon cooling, which indicates that the particle-water interfacial energy, $\sigma_{pw}$, also remains practically constant. One cannot expect also significant changes in $\sigma_{po}$ upon cooling as water-soluble surfactants are used in these experiments. Therefore, one cannot explain the particle detachment for chloromethyl latex in $C_{16}/C_{16}SorbEO_{20}$ emulsions using eq. (2).

Secondly, the formation of a rotator phase at the drop surface is expected to change strongly the energy of the three-phase contact line, because the interaction of the particle with the thin layer of rotator phase on the drop surface is energetically unfavorable. Indeed, the particle surface disturbs the molecular ordering in the layer of rotator phase, thus increasing the energy of the system in the zone of the particle-rotator phase contact. The latter effect can be analyzed thermodynamically using a modified Young-Laplace equation in which an additional term, accounting for the increased energy of the contact line, is introduced:

$$\cos\theta_\kappa = \frac{\sigma_{po} - \sigma_{pw}}{\sigma_{ow} - \kappa/r_c} \qquad (3)$$

Here $\kappa$ is the so-called "line tension" which is the excess energy per unit length of the contact line and $r_c$ is the radius of the contact line, Figure 7. As shown in the Appendix, eq (3) can be derived by minimizing the surface energy of the system with respect to the position of the particle at the fluid interface, under the assumption that all interfacial tensions and the line tension do not depend on the particle position (this is the regular procedure to derive the Young-Laplace equation). In our systems, we expect $\kappa \geq 0$, because the particle surface disturbs the molecular packing in the rotator phase which leads to an increase in the total energy of the system. Because the values of the line tension are typically small, the effect of line tension on the three-phase contact angle becomes important only for very small objects with micrometer and sub-micrometer contact lines, *e.g.* for heterogeneous nuclei in the process of new phase formation[53], micrometer sized oil lenses on the surface of surfactant solutions[54-56], small perforated holes in membranes or liquid films[57,58], and for micrometer and nanometer particles adsorbed on fluid interfaces[59]. Due to the small size of the latex particles in our systems and to the possible increase of the line tension with the formation of the rotator phase, this effect could be very significant for the emulsions studied here, as shown below.

For simplicity, in the following consideration we assume that the contact angle of the particles before the formation of rotator phase in the drop surface is not affected by the energy of the contact line and the term $\kappa/r_c$ is negligible in eq. (3). The respective initial contact angle is denoted as $\theta_i$ and is determined from eq. (2). The formation of a thin layer of rotator phase can increase significantly the magnitude of the line tension with concomitant decrease of the contact angle (for positive values of $\kappa$) even if the various interfacial tensions are assumed to remain (almost) unchanged. The changing contact angle is denoted as $\theta_\kappa$ and it should depend on the thickness of the surface layer of rotator phase, $h_{PL}$. A widely used approximation to estimate the value of $\kappa$ is to assume[57,58] that it is a product of the excess of interfacial energy in the contact zone and the width of the contact zone, *viz.* for the system under consideration $\kappa \approx h_{PL}(\sigma_{pr} - \sigma_{po})$, where $\sigma_{pr}$ is the interfacial energy at the contact zone of the particle with the rotator phase. Thus eq. (3) transforms into a transcendental equation for determination of the equilibrium contact angle $\theta(h_{PL})$. In the absence of rotator phase $h_{PL} = 0$, the contribution of the line tension is negligible, eq. (3) transforms into eq. (2), and $\theta_{\kappa \approx 0} = \theta_i$.

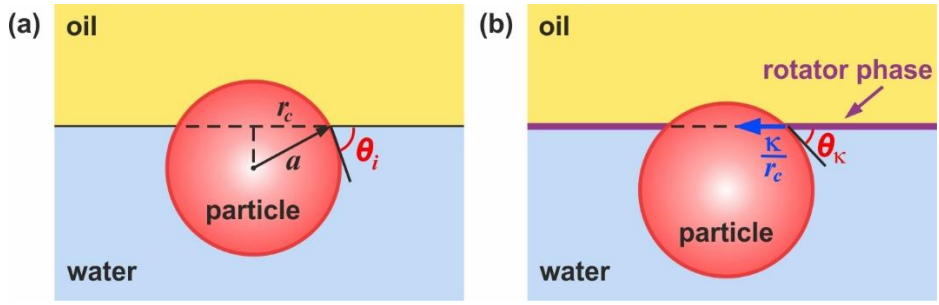

**Figure 7.** Schematic presentation of the three-phase contact angles $\theta_i$ (a) and $\theta_\kappa$ (b), and of the effect of the line tension, $\kappa$, on the equilibrium value of the contact angle. At sufficiently high positive values of $\kappa$, which may be caused by the formation of a surface layer of rotator phase, no equilibrium contact angle exists and the particle is ejected from the oil-water interface (see text).

We performed a theoretical analysis of the dependence $\theta(h_{PL})$ on the thickness of the rotator phase, see Appendix. As explained below, for realistic values of the various interfacial tensions, there is a well-defined thickness of the rotator phase at which the particle is ejected from the oil-water interface into the aqueous phase, because no energy minimum could be found for a particle residing on the oil-water interface. The physical meaning of this result is that the non-favored interaction of the particle surface with the rotator phase leads to significantly higher energy of the adsorbed particles and the system reduces its energy by particle desorption. In the Appendix we derive the following equation for the critical value of the dimensionless line tension which is sufficient to eject the particle from the drop surface:

$$\tilde{\kappa}_{cr} = \left(1 - \left(\cos\theta_i\right)^{2/3}\right)^{3/2} \qquad (4)$$

The numerical estimates showed that, for typical values of $\theta_i$ between *ca.* 20° and 60°, the dimensionless value $\tilde{\kappa}_{cr}$ varies between 0.01 and 0.23. Equation (4) can be transformed to show that the critical value of the dimensional line tension can be expressed as:

$$\kappa_{cr} = a\sigma_{ow}\left(1 - \left(\frac{\sigma_{po} - \sigma_{pw}}{\sigma_{ow}}\right)^{2/3}\right)^{3/2} \tag{5}$$

and it decreases with the decrease of: (1) particle size, (2) $\sigma_{ow}$, and (3) initial contact angle, $\theta_i$; see Figure A2 in Appendix.

Assuming that the dimensional value of $\kappa$ is proportional to the thickness of the rotator phase and to the difference in the interfacial tensions $\sigma_{pr}$ and $\sigma_{po}$, $\kappa \approx h_{PL}(\sigma_{pr} - \sigma_{po})$, we can now estimate whether the effect of line tension could be significant for the emulsions under study. Taking a realistic set of values: $a \approx 1$ μm, $\theta_i \approx 30°$, $\sigma_{ow} \approx 1$ mN/m, $h_{PL} \approx 10$ nm and $(\sigma_{pr} - \sigma_{po}) \approx 10$ mN/m, one estimates $\kappa \approx 1.0 \times 10^{-10}$ N/m and $\tilde{\kappa} \approx 0.1$, while the critical line tensions are $\kappa_{cr} \approx 0.28 \times 10^{-10}$ N/m and $\tilde{\kappa}_{cr} \approx 0.028$ for these values of the physical parameters. In other words, $\tilde{\kappa} > \tilde{\kappa}_{cr}$ and $\kappa > \kappa_{cr}$ which means that the particles would be ejected from the interface for such values of the line tension, caused by the perturbed layer of rotator phase. In contrast, in the absence of rotator phase or in the presence of a very thin layer of rotator phase, *e.g.* with $h_{PL} \approx 2$ nm (all other parameters remaining the same), $\kappa \approx 0.2 \times 10^{-10}$ N/m $< \kappa_{cr}$ and the particles have an equilibrium position at the interface with contact angle $\theta_\kappa < \theta_i$. Thus we see that upon increase of the thickness of the rotator phase, at an intermediate thickness of several nanometers (for the assumed realistic values of the physical parameters), the critical value of the line tension can be exceeded and, as a result, the particles will be ejected from the oil-water interface.

Summarizing the conclusions from this theoretical analysis, eq. (5) predicts that the particle desorption would occur in the process of emulsion cooling when a critical thickness of the surface layer of rotator phase is reached, and this critical thickness depends mostly on the initial contact angle, $\theta_i$, and on the values of $a$, $\sigma_{ow}$ and $(\sigma_{pr} - \sigma_{po})$. For line tensions and rotator phase thicknesses which are larger than the critical ones, no value of $\theta_\kappa$ could be found that would correspond to a stable equilibrium position of the particle on the fluid interface and the particles are ejected from the drop surface. If the critical thickness of the rotator phase and the respective critical line tension are not reached, then the particles remain adsorbed on the drop surface with contact angle, $\theta_\kappa$.

The above mechanism can be used to explain all the main trends observed in our experiments. For example, the particle detachment with purified $C_{16}$ is observed at slightly higher temperatures as compared to the systems with non-purified $C_{16}$, *cf*. Supplementary Figure S3 and S4. Under equivalent cooling conditions, the process begins at 18 °C with purified $C_{16}$ and at 17.1 °C with non-purified $C_{16}$ for 3 mM $C_{18}EO_{20}$ system, and at 19.6 °C for purified $C_{16}$ and at 18 °C for non-purified $C_{16}$ for 3 mM $C_{16}SorbEO_{20}$ system (all measured with 100 mM NaCl). One could propose the following explanation of these differences, supported by eq. (5). First, the surface-active impurities in the non-purified oil are expected to adsorb on the surface of the solid particle, thus reducing $\sigma_{po}$. As a result, the initial contact angle, $\theta_i$, should be larger for non-purified oil. Second, the same impurities are expected to adsorb in the region of the contact line as well, thus decreasing $\sigma_{pr}$. Under otherwise equivalent conditions, the surfactant adsorption is more pronounced at the interface with higher interfacial energy, $\sigma_{pr}$, and will thus decrease the difference ($\sigma_{pr} - \sigma_{po}$). This means that the impurities would lead to a lower value of $\kappa$ at the same thickness of the rotator phase. Respectively, a thicker layer of rotator phase would be needed to reach the critical value of the line tension needed for particle ejection, which requires lower cooling temperatures, just as observed experimentally.

To verify the mechanism proposed above, we performed additional experiments with purified hexadecane with 4.8 mM $C_{16}EO_2$ added as oil-soluble surfactant into the oil phase, along with the chloromethyl latex particles, 3 mM $C_{18}EO_{20}$ and 100 mM NaCl present in the aqueous phase. In this system, the interfacial energies, $\sigma_{po}$ and $\sigma_{pr}$, should be even lower due to the adsorption of the oil-soluble surfactant on the particle surface and in the three-phase contact zone, with related increase of $\theta_i$ and decrease of $\kappa$, under otherwise equivalent conditions. As predicted by eq. (5), despite the observed particle rearrangement process starting at ≈ 21.5 °C, no particle detachment was observed in this system. Thus, the added oil-soluble surfactant changed the conditions for ejection of the particles so that the critical line tension for particle ejection *via* the mechanism of "single particle detachment" explained above, was not reached during emulsion cooling (only some jammed particles were "ejected" from the drop surface *via* a different mechanism during the stage of intensive drop shape deformations, as explained below).

The above analysis could explain also why the detachment process was only observed with chloromethyl latex particles, whereas it was not observed with carboxyl and sulphate latex particles in the same oil-surfactant system. Most probably, the critical line tensions were not reached for the other types of particles, due to higher values of $\theta_i$ and/or lower values of ($\sigma_{pr} - \sigma_{po}$) which strongly depend on the type of particle surface.

The effect of surfactant concentration could also be explained by the same mechanism. Indeed, at 0.1 mM $C_{18}EO_{20}$ particle detachment was hardly observed, at 1 mM some of the particles detached from the drop surface but most of the particles remained adsorbed, while at 3 mM most of the particles detached upon cooling. The increase of surfactant concentration reduces the oil-water IFT, decreases the value of $\theta_i$ (thus decreasing the critical value of the line tension) and facilitates the formation of the rotator phase, thus increasing the actual line tension.

The temperature at which the particle detachment begins depends on the specific emulsion system and on the cooling rate applied. At higher cooling rates, the process begins at slightly lower temperatures, as compared to the slower cooling rate, which evidences that besides the absolute temperature needed for rotator phase formation, the time is also an important factor because longer times allow for development and completion of the rotator phase formation and for the concomitant increase of the line tension.

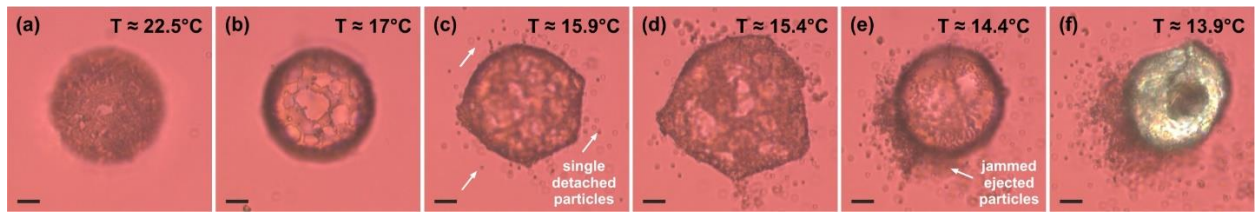

**Figure 8. Detachment of jammed particles.** (a) At high temperature, the particles are evenly distributed over the drop surface. (b) Rearranged layer of adsorbed particles observed at 17 °C. The rearrangement process started at $T \approx 21°C$. (c) Upon beginning of the drop shape transformation, few particles have detached from the drop surface. (d) Further drop shape evolution upon cooling. (e) Particle "ejection" upon return of the drop to spherical shape. The ejected particles are jammed and remain near the drop surface. (f) Drop after freezing. Many ejected particles are seen around the droplet. The system is: non-purified hexadecane droplet in 500 mM NaCl solution containing 1 mM $C_{18}EO_{20}$ surfactant, the adsorbed latex particles are with chloromethyl surface group. Cooling rate: 0.5 °C/min. Scale bars = 10 μm.

Besides the initial detachment of single particles observed around the beginning of the drop self-shaping process, a second type of particle "ejection" process was also observed, Figure 8. It takes place when the particle-particle interactions are strong and all particles are jammed together. The "ejection" of the jammed particles is observed in two different ways. In some of the experiments, the monolayer raft of adsorbed particles slides over the drop surface and most of the particles move outside the drop surface, while a small number of particles remain on the drop surface, see Supplementary Movie S3. Probably, this mechanism is a version of the mechanism for single particles, described above, with the only difference that the particles here are strongly attracted to each other (the electrostatic repulsion between them is completely suppressed) and are ejected as a raft. In other cases, the particle ejection occurs at a later stage, when the drops deform and occasionally return to spherical shape due to the action of interfacial tension. Upon such a return, the particle shell lags behind and gets detached from the deforming drop surface, see Supplementary Movie S4. Both types of "collective particle detachment" are observed in systems containing high electrolyte concentration, which suppress the electrostatic repulsion between the adsorbed particles, or when the electrolyte is $CaCl_2$.

These observations showed also that the particles adsorbed on the drop surface act as obstacles for the rotator phase formation. The observed processes of particle detachment reduce the system energy and allow for the subsequent formation of rotator phase and drop deformation in accordance with the general drop shape evolution scheme.

### *(b) Gorgon shape formation*

Another alternative process for drop surface deformation and expansion was observed for droplets which were highly covered with particles. Usually, these drops cannot change their shape upon cooling, because the particles suppress the regular formation of the plastic rotator phase. Instead, a fiber growth between the adsorbed particles was observed in these systems, Figure 9. In the experiments without particles, up to three fibers can be observed to grow from acute corners of a single shaped mother droplet. Here, the particles arrest the rotator phase formation locally, but fibers coated with rotator phase emerge from the gaps between the particles, and in this case the fiber number is dozens to hundreds. When the fibers are relatively thick, *i.e.* with diameter bigger than 1 μm, the oil drop with the numerous extruded fibers resembles a Gorgon's head with living snakes making up its hair, Figure 9d. Formation of such a shape has never been observed in the self-shaping droplets in the absence of adsorbed particles. The fact that instead of just one, two or three fibers, we observe formation of tens of fibers shows that the presence of adsorbed particles affects the rotator phase formation. Most probably, this is related to the fact that particles change locally the shape of the oil drop surface and make the local curvature suitable for the growth of fibers, similar to the case where they grow only from acute angles in flattened shapes.

The fiber growth starts at temperatures close to the temperature at which the drop shape transformations would normally start in the absence of particles and continue until the final drop freezing, see Supplementary Movie S5.

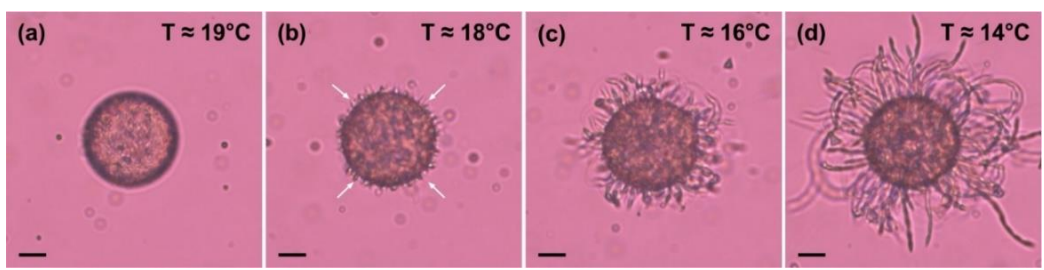

**Figure 9. Gorgon shape formation.** (a) At high temperature, the spherical drop does not have any asperities. Their growth begins around 18.9 °C. (b) The short asperities are well seen at 18 °C. (c-d) The asperities increase their length upon decreasing temperature. The system is: non-purified hexadecane drop stabilized by carboxyl latex particles, aqueous solution: 0.5 mM $CaCl_2$ and 1 mM $C_{18}EO_{20}$, cooling rate 1.2 °C/min. Scale bars = 10 μm.

The thickness of the extruded fibers varies depending on the alkane chain length, Figure 10. The fibers extruded from $C_{14}$ droplet are much thinner than those extruded from a $C_{15}$ droplet for the same surfactant-electrolyte combination, Figure 10a,b. The same difference was observed for the fibers formed from these droplets in the experiments without particles, which was hypothesized to be related to the molecular ordering at the drop surface and to the spontaneous curvature of the surface layer of rotator phase.[38] The fibers grown from a $C_{16}$ droplet in the presence of 1 mM $C_{18}EO_{20}$ and 0.5 mM $CaCl_2$ are even thicker. However, when NaCl is used as an electrolyte instead of $CaCl_2$, the fiber formation is much less pronounced. Even for droplets with very high particle coverage, usually only a few thin fibers are observed to grow. Therefore, for optimization of the process one should take into account not only the oil and surfactant in the system, but also the electrolyte used in the emulsion preparation.

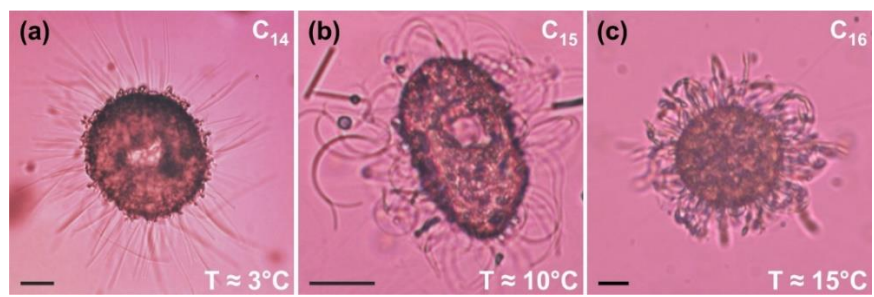

**Figure 10. Gorgon shape grown from alkane drops of different chain lengths with adsorbed carboxyl latex particles.** (a) Tetradecane drop in 100 mM NaCl and 1 mM $C_{18}EO_{20}$ solution; (b) Pentadecane drop in 100 mM NaCl and 1 mM $C_{18}EO_{20}$ solution; (c) Hexadecane drop in 0.5 mM $CaCl_2$ and 1 mM $C_{18}EO_{20}$ solution. Scale bars = 20 μm. Non-purified oils are used in the experiments.

### *(c) Drop shape transformations in the presence of adsorbed particles*

In the previous sections, we described the particle "detachment" mechanisms and growth of oil fibers between the adsorbed particles when the particle coverage is too high to allow the regular course of drop shape deformation. Here we discuss the particle behavior when the particles remain adsorbed on the drop surface upon the drop shape transformations.

The general drop shape evolutionary scheme observed for emulsion droplets without adsorbed particles[37,38] remains the same in the presence of adsorbed particles. However, as explained in the previous two sections, the adsorbed particles disturb the regular formation of a plastic rotator phase beneath the drop surface and usually the non-spherical shapes that form are much more irregular as compared to those observed without particles. Furthermore, the particle "ejection" process discussed above also shows that the adsorbed particles obstruct the self-shaping process. We note, that for all systems studied in the current paper, the observed interfacial tension at which the drop shape deformations begin is far above 0.01 mN/m which would allow one to explain the drop self-shaping process only with freezing of the surfactant adsorption monolayer without formation of plastic rotator phase multilayers.[41,42,44] Furthermore, the results presented in the current paper allow us to directly relate the interfacial tension measurements with the processes observed in the microscopy experiments. For $C_{16}$ alkane systems, the oil-water interfacial tension at which the drop deformations begin is estimated to be ≈ 6.3 mN/m for $C_{16}SorbEO_{20}$ system and ≈ 2.6 mN/m for $C_{18}EO_{20}$ system (if we extrapolate the measured values down to the lower temperature at which the deformations begin).

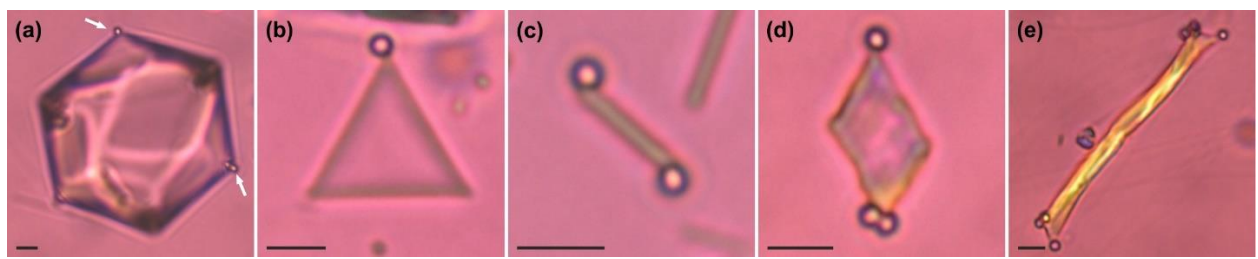

**Figure 11. Latex particles attached to the tips of non-spherical droplets.** When only few particles are adsorbed on the drop surface, they usually end up attached to the oil drop tips. (a-c) Fluid non-spherical droplets: (a) polyhedron; (b) triangular platelet; (c) rod-like particle. (d,e) Frozen non-spherical particles: (d) tetragonal platelet; (e) elongated tetragonal platelet. The systems are: chloromethyl latex particles adsorbed at hexadecane droplets dispersed in aqueous solution containing 100 mM NaCl and 1 mM surfactant: (a,e) $C_{18}EO_{20}$; (b,d) $C_{16}SorbEO_{20}$ and (c) $C_{18}SorbEO_{20}$. Scale bars = 5 μm.

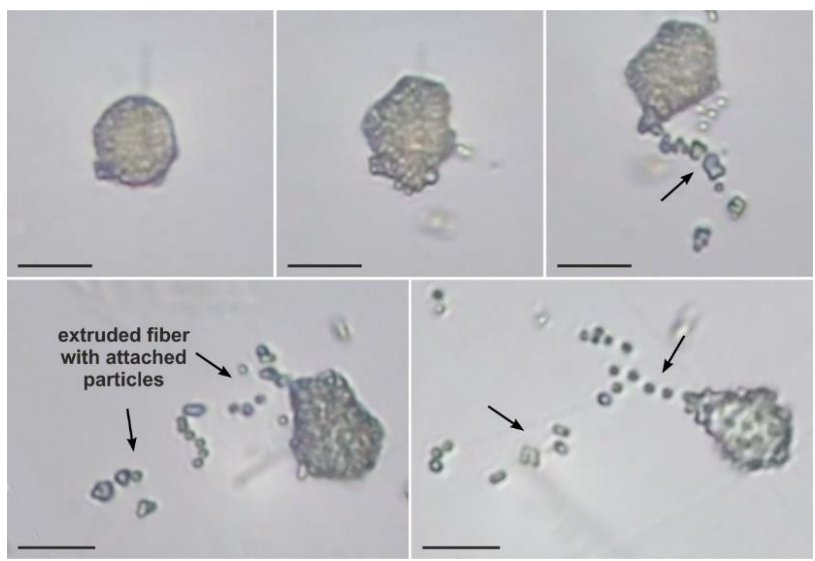

**Figure 12.** Fiber extrusion observed with a fully covered chloromethyl latex particle non-purified $C_{16}$ drop immersed in 1 mM $C_{18}EO_{20}$ solution containing 100 mM NaCl, cooling rate 0.3 °C/min. The drop deforms upon cooling and starts to extrude a thin fiber with attached particles on it (shown with arrows). Scale bars: 10 μm.

As the rotator phase formation was shown earlier to affect the attachment of particles to the surface, the presence of particles also affect the formation of rotator phase. In the presence of particles, the process of drop self-shaping is still observed, with the following differences. In the case when there are only few particles attached to the drop surface, upon cooling they end up attached to the defects[60,61], *i.e.* to the tips of the non-spherical shapes, Figure 11. This observation is in good agreement with the proposed mechanism about the particle rearrangement prior to the drop self-shaping, as the particles move initially toward areas with higher interfacial tension where the surfactant adsorption layer has not yet become frozen.

Another non-trivial particle behavior is observed when an oil droplet starts to extrude one or several thin fibers, see Figure 12 and Supplementary Movies S6 and S7. First, adsorbed particles situated near the extruding tip become attached to the fiber and move with it as it grows, leaving the droplet. In addition, particles that are adsorbed relatively far away from the extruding tip also become entrained and attached to the extruded fiber. This observation shows unambiguously that much of the material of the extruded fiber material comes from the drop surface rather than from the oil drop interior. Indeed, if the fiber material was coming entirely from the drop interior then there should not be a driving force pushing the adsorbed particles toward the extruding tip. The particle behavior can be easily understood, only if one assumes that the surface fiber material comes from the drop surface. Furthermore, this observation is also connected to the mechanism of fiber formation proposed previously.[38,45] As we explained in our previous work, the fibers should contain a rotator phase coat which enfolds the liquid interior in order to have the physical properties as observed in the experiments. Their ability to easily bend and further elongate after initial extrusion points to the presence of a liquid core, whereas their stability is much longer than expected for a liquid filament with respect to capillary principles, pointing out the presence of a rotator phase shell, rather than just a frozen monolayer. The "mother" drop interior is not filled with rotator phase, therefore fiber formation would if the surface material of the fibers comes from the drop surface where the rotator phase is situated. The observed particle behavior matches such a translation of the surface in the formation of the fibers.

**CONCLUSIONS**

We have expanded the field of particle desorption from Pickering emulsions with a new physical mechanism and shown conditions for the formation of a new type of complex hairy drops in the case of strongly adsorbed particles. Previous work in particle desorption has achieved it based on responsive particles which change their wetting properties[18-26] as well as external forces to remove the particles, including magnetic[23,24] and mechanical forces[30]. In the current paper we introduce a method for desorption of particles which takes advantage of *changes in the oil drop rather than in the particles*. Specifically, it is based on formation of a rotator phase close to the interface,[31-35] shown here for alkanes. Similar transformations occur in edible oils, such as triglycerides (the so-called α-phases[34]) and are even more prominent in the molecular packing of mixtures rather than pure compounds. Such interfacial transitions are the basis for self-shaping of emulsion droplets upon cooling,[35,37,38,45] including mixtures of compounds[62].

The current study explains how the drop self-shaping process is affected by the presence of adsorbed particles. The main conclusions can be summarized as follows:

(1) Prior to the freezing of the surfactant adsorption layer, the adsorbed latex particles are evenly distributed on the oil drop surface. After the adsorption layer freezes, the particles rearrange so that an ordered particle monolayer forms. This behavior is only observed in systems where the subsequent formation of a rotator phase is observed, *i.e*. when the drop self-shaping process is possible.

(2) The freezing of the surfactant adsorption layer, indicated clearly from the particle rearrangement, allowed us to unambiguously connect the measured interfacial tension from drop shape analysis to the interfacial tension present in the real emulsion experiments. This comparison showed that the drop shape deformations begin at $\sigma \approx 6.3$ mN/m for $C_{16}SorbEO_{20}$ and at $\sigma \approx 2.6$ mN/m for $C_{18}EO_{20}$, a value much larger than the one needed to explain the drop deformation with the formation of a frozen monolayer only. Therefore, a multilayer of plastic rotator phase should be present to allow the observed deformations in the systems studied by us.

(3) A single-particle detachment process is observed upon the beginning of rotator phase formation for some of the systems studied. It is explained by a decrease of the three-phase contact angle, due to the formation of a plastic rotator phase on the drop surface. As a result, the conditions for particle attachment change and the particles are ejected from the drop surface. In some systems, the particles are jammed and are "ejected" from the drop surface together as a shell.

(4) A theoretical model of the particle detachment is developed which considers explicitly the effect of the line tension (excess energy of the three-phase contact line) on the equilibrium position of the particle on the oil-water interface. This model shows that realistic positive values of the line tension could result in the absence of an equilibrium particle position on the fluid interface for particles which otherwise (in the absence of line tension) would have a finite equilibrium contact angle.

(5) The shape sequence in the general drop shape evolutionary scheme is not affected by the presence of adsorbed particles, though the observed shapes are much more irregular than those in absence of particles. As the particles adsorbed at the drop surface are observed to be "dragged" toward the extrusion tip, much of the material for fiber extrusion comes from the surface of the oil droplet.

(6) A new type of "Gorgon"-like drop shape is observed when the particles are jammed on the drop surface and the particle coverage is high. Tens of fibers grow outwards from the drop surface in this case.

Our demonstration that the particle desorption and self-shaping are driven by the same mechanisms implies straightforward generalization of these experiments. We have shown the self-shaping process in other oil types such as long chain alcohols, triglycerides, other alkanes and mixtures of the above with oils that don't self-shape or exhibit rotator phases alone, therefore such transformations are likely to display the similar particle desorption behavior demonstrated in this work.

The model including the line tension effects, developed in this paper to explain particle ejection from the oil-water interface, could find applications in the quantitative description of other systems, *e.g.* for micro- and nanoparticle-stabilized Pickering emulsions and for the interfacial rheology of particle-laden interfaces. Note that the line tension effects are inversely proportional to the particle size which predicts that they will be even more pronounced for sub-micrometer and nanometer particles.

**Acknowledgements:** The beginning of this study was partially funded by the European Research Council grant EMATTER (# 280078). Authors are grateful to Mrs. Mila Temelska (Sofia University) for performing the measurements of the interfacial tensions. The authors thank to Dr. Stoyan Smoukov (Queen Mary University of London, UK) for purchasing some of the latex samples used in the study (through EMATTER project # 280078) and for his suggestion to study the effect of silica particles on the self-shaping phenomena – a preliminary study which turned out to be unsuccessful. The current study falls under the umbrella of European Network COST CA 17120.

**Authors contributions:**

B.P.B suggested studying the effect of adsorbed particles on the self-shaping phenomenon; S.C. planned and designed the experimental study (with input from B.P.B. and N.D.); J.V. and D.C. performed the experiments and summarized the results; S.C. and D.C. analyzed the results, interpreted them to reveal the various mechanisms described and made theoretical calculations for the line tension effect; N.D. suggested and clarified the role of the line tension in the particle desorption process and suggested the mechanism for Gorgon drops formation; D.C. wrote the first draft of the manuscript, prepared the figures, movies, Appendix, reference list and Supplementary Information; N.D. prepared the final version of the manuscript. All authors participated in the discussions and critically read the manuscript.

# Appendix

# Effect of line tension on the three-phase contact angle at the particle surface

The energy of the system in its initial state, when the particle is fully immersed in the aqueous phase (Figure A1a) can be expressed as:

$$E_0 = A_{pw}\sigma_{pw} + A_{ow}\sigma_{ow} = 4\pi a^2 \sigma_{pw} + \pi a^2 \sin^2\theta \sigma_{ow}, \quad (A1)$$

where $a$ is the particle radius, $\theta$ is the three-phase contact angle, $\sigma$ is the interfacial tension, $A$ is the interfacial area and subscripts "$p$, $w$, $o$" denote particle, water and oil phase respectively. After the particle adsorption on the interface, the energy of the system becomes:

$$E = A_{pw}\sigma_{pw} + A_{po}\sigma_{po} + 2\pi\kappa r_c = 2\pi a^2 (1+\cos\theta)\sigma_{pw} + 2\pi a^2 (1-\cos\theta)\sigma_{po} + 2\pi\kappa a \sin\theta, \quad (A2)$$

where $\kappa$ is the line tension acting on the three-phase contact line and $r_c = a\sin\theta$ is the radius of the three-phase contact line, see Figure A1b and Figure 7.

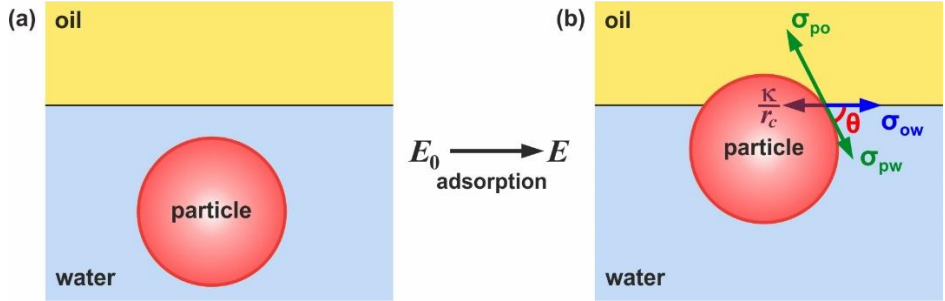

**Figure A1.** Schematic representation of the energy of the system before (a) and after (b) the adsorption of a particle on the oil-water interface.

The energy difference between the adsorbed and the initial states is:

$$\Delta E = E - E_0 = 2\pi a^2 (1-\cos\theta)(\sigma_{po} - \sigma_{pw}) - \pi a^2 \sin^2\theta \sigma_{ow} + 2\pi a\kappa \sin\theta. \quad (A3)$$

For negligible line tension, the adsorbed particles have three-phase contact angle, $\theta_i$, which can be expressed by the conventional Young-Laplace equation:

$$\sigma_{po} - \sigma_{pw} = \sigma_{ow}\cos\theta_i \quad (A4)$$

Next, we substitute eq. (A4) into eq. (A3) to obtain:

$$\Delta E = 2\pi a^2 (1-\cos\theta)\sigma_{ow}\cos\theta_i - \pi a^2 \sin^2\theta\sigma_{ow} + 2\pi a\kappa\sin\theta \tag{A5}$$

$$\frac{\Delta E}{\pi a^2 \sigma_{ow}} = 2(1-\cos\theta)\cos\theta_i - \sin^2\theta + 2\sin\theta\frac{\kappa}{a\sigma_{ow}} \tag{A6}$$

We define a dimensionless energy difference and dimensionless line tension as follows:

$$\Delta\tilde{E} = \frac{\Delta E}{\pi a^2 \sigma_{ow}} = 2(1-\cos\theta)\cos\theta_i - \sin^2\theta + 2\tilde{\kappa}\sin\theta \tag{A7}$$

$$\tilde{\kappa} = \frac{\kappa}{a\sigma_{ow}} \tag{A8}$$

To find the equilibrium position of the particle on the oil-water interface in the presence of a noticeable line tension, we can use the derivative of the adsorption energy difference with respect to the contact angle, $\theta$, at fixed values of the interfacial and line tensions (meaning fixed $\theta_i$ as well):

$$\frac{d\Delta\tilde{E}}{d\theta} = 2\sin\theta(\cos\theta_i - \cos\theta) + 2\tilde{\kappa}\cos\theta \tag{A9}$$

This function has a critical point (maximum, minimum or inflection point) if:

$$\frac{d\Delta\tilde{E}}{d\theta} = 0 \;\Rightarrow\; 2\sin\theta_\kappa(\cos\theta_i - \cos\theta_\kappa) + 2\tilde{\kappa}\cos\theta_\kappa = 0 \;\Leftrightarrow\; \tilde{\kappa} = \sin\theta_\kappa\left(1-\frac{\cos\theta_i}{\cos\theta_\kappa}\right), \tag{A10}$$

where $\theta_\kappa$ is the three-phase contact angle for which the function $\Delta\tilde{E}(\theta,\theta_i,\tilde{\kappa})$ has a critical point. Note that for $\tilde{\kappa} > 0$ (as supposed in our consideration) $\theta_\kappa < \theta_i$.

If the dimensional parameters are used, eq. (A10) takes the form:

$$\cos\theta_\kappa = \frac{\sigma_{po} - \sigma_{pw}}{\sigma_{ow} - \kappa/r_c} \tag{A11}$$

which is equivalent to eq. (3) in the main text.

Figure A2a shows the plot of the function $\tilde{\kappa}(\theta_\kappa)$ for three different values of $\theta_i$. Remarkably, the function $\tilde{\kappa}(\theta_\kappa)$ has a maximum which is denoted by $\tilde{\kappa}_{cr}$, Figure A2b. This means that for any value of $\tilde{\kappa} > \tilde{\kappa}_{cr}$, the energy difference function (equation A10) has no minimum or maximum for any value of $\theta_\kappa$. For the limiting value, $\tilde{\kappa} = \tilde{\kappa}_{cr}$, the energy difference function has an inflection point at a certain value of $\theta_\kappa = \theta_{cr} \geq 0$ (again no minimum exists in this case). Only for $\tilde{\kappa} < \tilde{\kappa}_{cr}$ does the energy function have a minimum which corresponds to an equilibrium position of the particle with contact angle $\theta_\kappa < \theta_i$, see Figure A2c.

Finally, we determine the position and the value of the maximum of $\tilde{\kappa}(\theta_\kappa)$:

$$\frac{d\tilde{\kappa}}{d\theta_\kappa} = \cos\theta_\kappa - \cos\theta_i \frac{1}{\cos^2\theta_\kappa} \tag{S12}$$

$$\frac{d\tilde{\kappa}}{d\theta_\kappa} = 0 \implies \theta_\kappa = \arccos\left(\sqrt[3]{\cos\theta_i}\right) \tag{A13}$$

$$\tilde{\kappa}_{cr} = \sin\left(\arccos\left(\sqrt[3]{\cos\theta_i}\right)\right)\left(1-\left(\cos\theta_i\right)^{2/3}\right) = \left(1-\left(\cos\theta_i\right)^{2/3}\right)^{3/2}, \tag{A14}$$

where we have used the trigonometrical identity $\sin\left(\arccos\left(\sqrt[3]{\cos\theta_i}\right)\right) = \sqrt{1-\left(\sqrt[3]{\cos\theta_i}\right)^2}$.

If we substitute back the dimensionless parameters, eq. (A14) becomes:

$$\kappa_{cr} = a\sigma_{ow}\left(1-\left(\frac{\sigma_{po}-\sigma_{pw}}{\sigma_{ow}}\right)^{2/3}\right)^{3/2} \tag{A15}$$

$$\kappa_{cr} = a\sigma_{ow}\left(1-\left(\cos\theta_i\right)^{2/3}\right)^{3/2} = h_{PL}\left(\sigma_{pr}-\sigma_{po}\right) \tag{A16}$$

Thus we have derived explicit equations for the line tension effect on all important quantities characterizing the particle position on the oil-water interface.

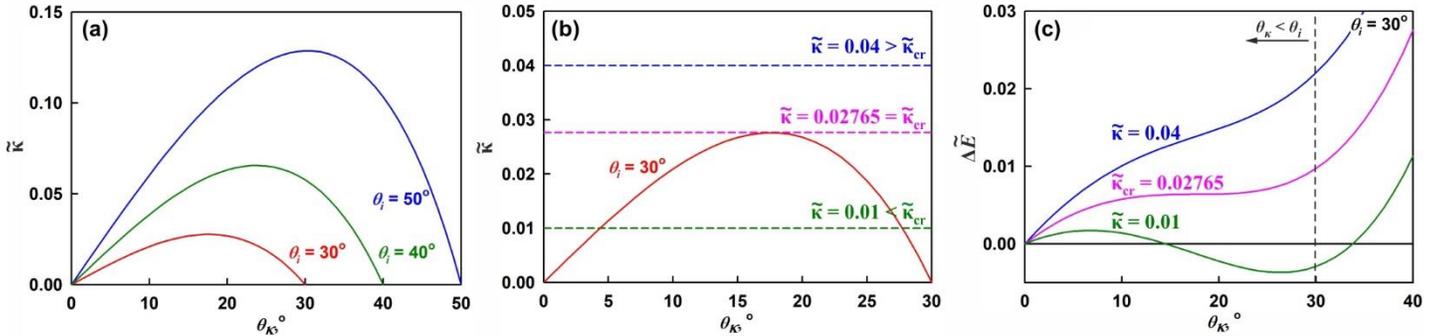

**Figure A2.** (a) Plot of the function $\tilde{\kappa}(\theta_\kappa)$ for three different values of the initial three-phase contact angle $\theta_i$. (b) Function $\tilde{\kappa}(\theta_\kappa)$ plotted for $\theta_i = 30°$ with three reference lines showing different values of $\tilde{\kappa}$ for which $\Delta\tilde{E}$ is plotted in (c). (c) Function $\Delta\tilde{E}(\theta_\kappa)$ plotted for three different values of $\tilde{\kappa}$. This function has a minimum only when $\tilde{\kappa} < \tilde{\kappa}_{cr}$ and $\tilde{\kappa}_{cr} \approx 0.02765$ for $\theta_i = 30°$.

# **Graphical abstract**

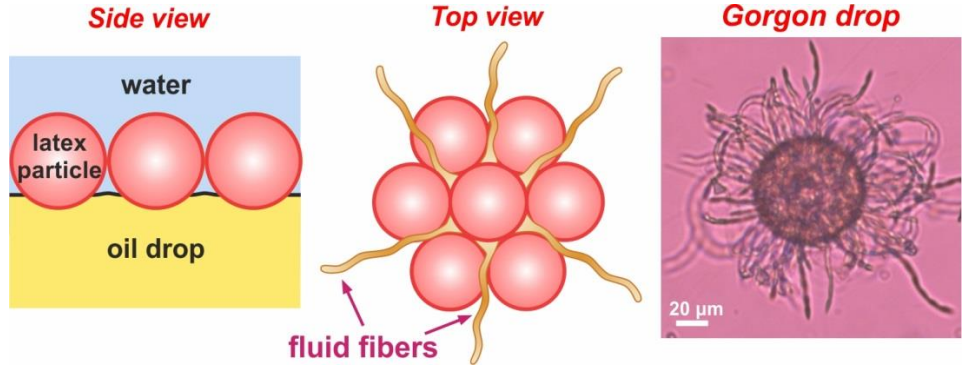